# Experimental determination of the local optical conductivity of a semiconducting carbon nanotube and its modification at individual defects

**Authors:** Ryosuke Senga[1]*, Thomas Pichler[2], Yohei Yomogida[1], Takeshi Tanaka[1], Hiromichi Kataura[1] and Kazu Suenaga[1]*

**Affiliations:**

[1]*Nano-Materials Research Institute, National Institute of Advanced Industrial Science and Technology (AIST), Tsukuba 305-8565, Japan.*

[2]*Faculty of Physics, University of Vienna, Strudlhofgasse 4, A-1090 Vienna, Austria*

*Correspondence to: Ryosuke Senga (Ryosuke-senga@aist.go.jp) and Kazu Suenaga (suenaga-kazu@aist.go.jp)

**Abstract:** Measurements of optical properties at nanometre-level are of central importance for characterization of optoelectronic device. It was, however, hardly possible for the conventional light-probe measurements to determine the local optical properties from single quantum object with nanometrical inhomogeneity. Here we demonstrate the first successful determination of the absolute optical constants, including the optical conductivity and absorption coefficient, for an individual carbon nanotube with defects by comparing energy loss spectroscopy and optical absorption. The optical conductivity obtained from a certain type of defects indeed presents a characteristic modification near the lowest excitation peak ($E_{11}$) where excitons and non-radiative transitions as well as phonon-coupled excitations are strongly involved. Detailed line-shape analysis of the $E_{11}$ peak clearly shows different degree of exciton lifetime shortening and electronic state modification which is variable to the defect type.

**Introduction**

Optical properties of low-dimensional materials are strongly governed by exciton behaviour in which bound electron-hole pairs are confined in the limited dimensionality. Especially excitons generated in a semiconducting carbon nanotube, which is an ideal one-dimensional system, exhibit an extremely strong binding energy (a few 100 meV[1,2]) and therefore show unique behaviours such as a large exciton Bohr radius (a few nm[3]), an unusual stability at room temperature and a long diffusion distance (more than 500 nm[4,5]). In general, such stable and free mobile exciton behaviours as well as other fascinating physical properties of carbon nanotubes should be advantageous for the light-emitting devices or photovoltaic batteries. However the luminescence quantum yields of carbon nanotubes is not typically high (a few percent[6–8]). The origin of the low luminescence quantum yield has been believed as the non-radiative exciton decays happened at non-periodic structures such as defects, edges or attached impurities[9,10]. Meanwhile, the enhancement of emissions in carbon nanotubes presumably caused by the site-specific exciton recombination at such defects or chemically doped sites has been also reported[11–13]. Therefore, the direct correlation between the local exciton behaviour and the atomically identified defects must be addressed for further understandings. However, it is hardly possible for conventional measurements using light probes to directly correlate the local optical properties to the atomic structures because of their spatial resolution limited by the optical wavelength. Although the near-field scanning optical microscopy can overcome the diffraction limit of the optical source, its spatial resolution is a few tens nm[14] and still difficult to probe single defects. Hence, these tip enhanced near-field results are complicated by the exact knowledge of the unknown joined density of states (DOS) at the defect, the extension of wave function of the exciton and selection rules which are modified by the metal tip.

On the other hand, an electron probe in a transmission electron microscope (TEM) which has a much smaller wavelength ($\lambda$=0.05 Å at 60 kV) can picture the atomic structures of materials with an atomic resolution and also induce the optical excitations in the single quantum objects. Electron energy-loss spectroscopy (EELS) which is often combined with a TEM can quantify the loss energy for the electron excitations across the Fermi level in materials over a wide energy range. In addition, the recent development of the monochromator for TEM has pushed up the EELS energy resolution better than a few 10 meV and allow us to identify the absorption peaks for optical excitations including the optical band gaps in an ultraviolet, visible and near infrared energy range in the various semiconducting materials such as silicon[15], 2D transition metal dichalcogenides[16], perovskite nanoparticles[17] and single-walled carbon nanotubes (SWNTs)[18–20]. In addition, the wavelength of incident electron beam which is much smaller than the single quantum object size provides a near-field condition and therefore enables us to excite the extra transitions which are basically forbidden in the light probe measurements.

However, despite of these advantages of achievable higher spatial resolution and sensitivity, the usage of local EELS for optical properties is still limited because of the lack of studies that quantitatively links the local EEL spectra to the optical properties. In fact, the local optical bandgaps of several materials measured by EELS have shown slightly higher values than those measured by optical absorption spectroscopy (OAS), though the difference has not been fully discussed and sometimes reluctantly concluded as "almost same". In addition, only few EELS studies present the absolute optical constants calculated from EEL spectra and none of them treats the single quantum objects nor the individual defects.

Here we demonstrate the absolute optical constant measurements of single isolated SWNTs through the Kramers-Kronig (KK) transformation of the local loss function obtained by using the monochromated TEM. The absolute dielectric function for the single isolated SWNT was successfully obtained for the first time by directly comparing the KK transformed EELS with OA spectra collected from the same chirality SWNTs prepared by the high-yield chirality separation method[21,22]. Furthermore, the local modifications of the exciton behaviours were detected and their lifetime shortening and extra states of excitons were unambiguously discriminated at specific defective sites on a SWNT.

**KK analysis to combine EELS and OAS**

The optical and electron pathways in OAS and EELS have some parts in common (Fig. 1a) and their spectra are analogous but not completely equivalent (Fig. 1b). The OA spectrum in Fig. 1b was obtained from a suspension of single chirality for (9, 2) SWNTs prepared by the high-yield chirality separation method using a gel column chromatography[21]. Meanwhile, the EEL spectrum in Fig. 1b was taken from a single isolated (9, 2) SWNT (freestanding in vacuum). The chiral index of the target SWNT was confirmed by TEM images (inset). Since the both OA and EEL spectra in Fig. 1b are reflecting the pure (9, 2) SWNT properties, they should be directly compered in a complex dielectric function $\varepsilon = \varepsilon_1 + i\varepsilon_2$. According to the dielectric theory, OA and EEL spectra are roughly proportional to $\varepsilon_2$ and to the loss function $\varepsilon_2/(\varepsilon_1^2+\varepsilon_2^2)$, respectively, and show comparable (but not completely same) features in a diluted system (Fig. S1). In fact, the both spectra show common features including several sharp peaks regarding to the excitons (and the interband transitions) from van Hove singularities in the valence band to ones in the conduction band. But the first peak ($E_{11}$) in EELS is 200 meV higher than that in OAS (Fig. 1b). The gap in peak positions between EELS and OAS is more prominent in the higher order absorption peaks ($E_{22}$ to $E_{55}$). Note that the EEL spectrum has intermediate peaks indicated by black arrows between $E_{11}$, $E_{22}$ and $E_{33}$ which are basically silent in the OA spectrum (Fig. 1). Since the electron beam in our experimental condition has a much smaller wavelength than the tube diameter and also integrates a large $q$-momentum space, transitions induced by a wide range of polarizations including both parallel and perpendicular to the tube axis are allowed. Therefore the intermediate peaks can be assigned as $E_{12}$ (+$E_{21}$) and $E_{23}$ (+$E_{32}$) which are observable only when the perpendicular polarized light comes into CNTs in the case of OAS[23]. Indeed, when the parallel electron beam passes by an aloof position, more than 10 nm apart from the nanotube, these $E_{12}$ and $E_{23}$ become silent (Fig. S2b).

In order to obtain the absolute dielectric function, the Re[ε]=$ε_1$ can be derived from the EEL spectrum by a Kramers-Kronig (KK) relation. For the loss function the KK relation is:

$$\mathrm{Re}\left[\frac{1}{\varepsilon(q,\omega)}\right] - 1 = \frac{1}{\pi} P \int_{-\infty}^{\infty} \frac{d\omega'}{\omega' - \omega} \mathrm{Im}\left[\frac{1}{\varepsilon(q,\omega)}\right]$$

where $P$ denotes the Cauchy principle part of the integral, avoiding the pole at $\omega=\omega'$. Basically ε depends on a frequency $\omega$ and a wave-vector $q$. Our experiments were performed in different $q$ conditions between EELS and OAS as the converged electron beam ($q \neq 0$) was used for EELS, while the light source in OAS was a parallel beam ($q = 0$). Since the interband excitations between the van Hove singularities in semiconducting SWNTs have vanishingly small $q$ dependence[24] and the plasmon of the individual tubes have a strong forward scattering component we can still directly compare EELS and OAS by simply discussing the energy (frequency) $\omega$ dependence of ε($\omega$) in the following analysis. The further processing of EEL spectra before the KK transformation including the background subtractions is described in Figs. S3, S4 and Supplementary text.

For the KK transformation, the static dielectric function $\varepsilon_\infty = Re[\varepsilon(0)]$ which is the intercept of $ε_1$ must be known as input for normalizing the loss function. However, the static dielectric function $\varepsilon_\infty$ for an individual SWNT has never been experimentally obtained. Therefore, we have performed an iteration analysis by shifting the value $\varepsilon_\infty$ until the calculated $\varepsilon_2$ from EELS fits to the OA spectra. During the process, the normalized EELS spectra (absolute Im[-1/ε]) are also derived by the KK sum rule as shown in Fig. 2a.

**Absolute optical constants for individual carbon nanotubes**

Figure 2 shows the resultant dielectric function $\varepsilon_1$ and $\varepsilon_2$ and optical constants (refractive index n=$\sqrt{\varepsilon_1 + |\varepsilon|}/\sqrt{2}$ and extinction coefficient κ=$\sqrt{-\varepsilon_1 + |\varepsilon|}/\sqrt{2}$) derived by the KK transformation at $\varepsilon_\infty$=2.16. For this value the absorption coefficient α=4πk/λ, where λ denotes the wavelength, calculated from the resultant ε shows an excellent agreement with the OA spectra in their peak positions (Fig. 2b). In this analysis, the experimental broadening in the EEL spectra must be taken into account in comparison to optics system which does not need to consider the experimental broadening. Indeed, a 30 meV Gaussian broadening in the full width at half-maximum (FWHM) to the OA spectrum (Fig. S2), which is reasonable for the energy resolution of the monochromated electron source (*dE*=30~50 meV in FWHM), well fits to the left side of the first peak in the optical absorption coefficient as well as the optical conductivity calculated from EEL spectra. However, the right sides of the first peak in Fig. 2c and 2d are affected by the exciton-phonon coupling and require additional line fittings as described later. Note that the same experimental broadening is not applicable for the higher order peaks ($E_{22}$~) because of the slight defocus on the higher energy peaks in EELS spectrometer.

The absorption coefficients can be directly compared with the OA spectrum (normally presented as an absorbance), since an absorbance *A* is proportional to the α based on Lambert-Beer's low: *A*=α*LC*, where *L* and *C* denote the light pass length and the sample concentration, respectively. From the calculated absorption coefficient, one can roughly estimate the number of SWNTs contributed to the OA spectra. In our experiments, more than 30 SWNTs could be contributed to the OA spectra.

The static dielectric function obtained in the analysis ($\varepsilon_\infty$ =2.16) is a characteristic value for the single semiconducting SWNT and applicable for other chirality semiconducting SWNTs (Fig. S5). Interestingly, the value is much lower than the theoretically predicted one ($\varepsilon_\infty$ =30[25]). In order to crosscheck the sum rule we also compared the total area under the loss function with other molecular nanostructures like $C_{60}$ crystal. Although in the latter case we have a bulk plasmon we can compare the effective electron density in the two systems. We observe that for an individual SWNT the electron density is about 50% of those in $C_{60}$ which allows us to estimate the effective extension of the electron wave function in the SWNT. We observe an effective tube radius of about 6.8 Å. This means that the electron wave function extends about 2.8

Å over the observed structural radius into free space which is in good agreement with the observation that the loss function is strongly influenced by another tube if their distance is closer than 1 nm[19].

**Detailed excitonic behaviours derived from the fine structure analysis**

The real part of the optical conductivity $Re[\sigma] = -\varepsilon_0 \omega \varepsilon_2$, where $\varepsilon_0$ denotes the permittivity of vacuum, is also an important function to investigate the optical properties and to compare with theory (Fig. 2c). Especially an $E_{11}$ transition directly reflects the exciton properties such as the oscillator strength and the excitons lifetime which can be estimated from the absolute intensity and line-widths of the exciton peaks in $Re[\sigma]$. An $E_{11}$ transition consists of three components as shown in Fig. 3a: I) an exciton which is asymmetrically modified by the joined DOS of the SWNTs which is also observed in OAS and in the far field EELS (Fig. S2b), II) an exciton-phonon coupling and III) a non-radiative band-to-band transition. The fitting by the standard oscillator model (Kramers-Heisenberg dielectric function) of the $E_{11}$ transition is the simplest way to roughly investigate the exciton properties (Fig. S6 and Supplementary text). However the model cannot fully explain the obtained results because of the asymmetric shape of component I. Thus, we performed a further line-shape analysis including a scaled joined DOS for the first exciton peak (component I) by employing the model function made of a reciprocal square root singularity multiplied by a Gaussian cumulative and a Lorentzian function. The function allows us to simulate the experimentally broadened joined DOS by the Gaussian cumulative component and extract the relative intensities and the damping factors from the FWHM of the Lorentzian component.

Component II has multiple damping peaks which are equidistantly separated between 32 and 37 meV. This is fully consistent with the damped Franck Condon satellites (FCS) reflecting the coupling between the exciton and the phonon corresponding to the radial breathing mode (RBM)[26,27]. These FCS are basically observed in molecules and are usually strongly damped in solids[28] and therefore could be seen as a direct proof for the "molecular" behaviour of the local dielectric function of a single SWNT. Note that the exciton-phonon related to G-band bound state at about 200 meV above to the main peak in OAS (Fig. 2b), which is also reported in previous reports[26,27,29], cannot be clearly found in EEL spectra. This is probably because the exciton-RBM phonon coupling is more prominent in our near-field condition and buries the components of exciton-phonon related to G-band coupling.

The energy gap between the peak in component III (indicated by $E_{11}$') and the exciton peak ($E_{11}$) is approximately 330 meV, which is comparable to the exciton binding energy for similar size nanotubes[2,30]. In addition, the gap between $E_{11}$ and $E_{11}$' is almost inversely proportional to the tube diameter (Fig. 3c). Note that the energy gap between $E_{11}$ and $E_{11}$' indicated by the closed squares in Fig. 3c were measured from the original EEL spectra in our previous work[19]. It shows slightly lower values but the same tendency as the exciton binding energy measured by two-photon excitation spectroscopy (opened triangles in Fig. 3c plotted from ref. 30). The $E_{11}$-$E_{11}$' gap measured from the optical conductivity (opened circles) presents more consistent values with the two-photon excitation spectroscopic results. Therefore, one can reasonably assign the peak as the direct band-to-band transition (non-radiative transition). Basically, such direct band-to-band transitions cannot be seen in OAS. However our experimental condition, a large $q$ momentum integrated near-field probe, in which multipolar-induced transitions are allowed as well as dipolar-induced ones, can excite any possible interband transitions including the direct band-to-band transition at higher momentum transfer[31]. Inversely, when we collect the loss spectra with a high $q$-resolution ($q \cong 0$) electron beam at an aloof position in which only dipolar induced transitions are allowed, the band to band transition as well as $E_{ij}$ transition peaks ($i \neq j$) become silent (Fig. S2b). Component II and III are fitted by Voigt functions in which a Gaussian line width is fixed at 30 meV to include the experimental broadening effect.

**Locally modulated optical property at defects**

Figure 4 presents local modifications in the optical conductivity corresponding to the atomic structure variations. Fig. 4a shows a 230 nm long (9, 2) SWNT suspended between TEM grid with three different defects (namely, defect B, C and D). We already showed in our previous paper that the exciton in SWNTs is well localized and its spatial extension is limited to 3 ~ 5 nm at most[19]. The three defects in the (9, 2) SWNT shown in Fig. 4 are apart from each other more than 50 nm and therefore each defect does not interfere with the others. Figure 4e shows the optical conductivity at different defect structures (B, C, and D) after the same treatment described above. At a first glance, one can find that only the $E_{11}$ features are completely distinctive corresponding to the defect types while the other higher order peaks ($E_{22}$~) do not show obvious changes. If a charge transfers to the defects or attached impurities, it is reasonable that only the $E_{11}$, which is quite sensitive to the Fermi level, is affected. Such hole-doping effects have been reported in functionalized SWNTs[13,32].

In addition, we can picture more detailed modifications in the features near $E_{11}$ transition corresponding to the defect type by the line shape analysis described above

(Figs. 4b-4d). Extracted line widths (lifetime) of the exciton peak and integrated intensities for each component are listed in Table 1. The relative intensity compared to the defect-free region ($I/I_0$) are also shown in parentheses.

Defect B which is a small bump on the nanotube wall shows only small exciton lifetime reduction from the defect-free region. The integrated intensity for the exciton decreases to 86%, while the area under the non-radiative band-to-band transition (component III) increases by 35 %. In addition, the non-radiative peak slightly shifts probably due to the band structure modification. Defect B also has a small pre-peak at about 50 meV below the $E_{11}$ main peak (indicated by black arrow). The energy position of the pre-peak is close to the dark-exciton states which can become "active" or "bright" when the symmetry breaks by the external field or existence of non-periodic structures[12,33].

Defect C consists of multiple defects with a complex structure, which may involve a fullerene-type defect[34], and hardly shows a noticeable lifetime changes in the radiative component. The integrated intensity of component I largely decreases to 61% at this defect-type. In addition, component II and III are reduced here, too. The shift of the non-radiative component ~0.07eV is also prominent, which suggests a defect state generated between the intact bandgap or a decreased exciton binding energy.

The lifetime of exciton at the defect D where a few nm sized carbon impurity is attached is extremely shortened down to almost one-seventh. The integrated intensity of component I is also reduced by 40%. However, defect D has an extra peak below the $E_{11}$ main peak (indicated by the black arrow in Fig. 4d). The extra pre-peak can be assigned as a brightened inactive exciton state induced by the symmetry breaking as is the case with defect B. Another interpretation for the pre-peak is a contribution from charged excitons (trions). Since holes seem to be induced effectively at defect D as seen from the reduction of main $E_{11}$ peak, trions may be reasonably generated[13].

The contribution from the FCS is enhanced in all the defects i.e. the main $E_{11}$ exciton peak has a lower spectral weight as compared to the $I_{FCS}$. This means the exciton localized at defects can behave more likely as molecules. On the other hands, the lifetime of the exciton at defect B and defect C stays the same as one for the defect-free region. This is fully explainable by the fact that the exciton wave function is more localized at the defect but the distortion of the wave function is not strong enough to influence the exciton lifetime. Only the wave function at defect D has a strong distortion to significantly influence the exciton lifetime. The result suggests how important the exact knowledge of the defect type is in order to perceive the exciton behaviours.

**Discussion**

Since EELS investigates the excitation process but not emission process, it is hardly possible to directly know the exact emission properties of individual defects such as emission quantum yields. Nevertheless, our experimental results suggest possible scenarios for exciton properties at defects from the following key factors determining the absorption and emission properties.

First, the probability of exciton trapping at defects can be simply led by the exciton lifetime. Because a typical diffusion length of exciton is estimated at the order of a few hundred nanometers, many of the exciton created on the other part of SWNT can travel along substantially before they annihilate. In this viewpoint, defects with shorter exciton lifetime can be a trap site of travelling excitons. For example, in Fig. 4, defect D can gather more excitons while defect B and C cannot be hot spots for the exciton recombination.

The second factor is the existence of additionally formed optically active states in the DOS. The several previous reports demonstrated the improved quantum emission yields induced by the slightly red-shifted emissions in addition to the initial $E_{11}$ emission[11–13]. Such locally formed additional active states have certain oscillator strengths and should be detectable as absorption peaks in our near-field experimental condition which integrates a whole momentum space. Indeed, defect B and D has pre-peaks below the $E_{11}$ exciton peak while defect C shows no pre-peaks.

The third factor is Fermi level shifts induced by the hole-doping. Basically a hole-doped energy state serves as an emission quenching site because the excited state can be likely relaxed through the Auger dissociation process[10]. Since all the defects in Fig. 4 show the reduction of $E_{11}$ intensity, a certain exciton quenching contribution by Auger dissociation should be taken into account.

In addition, the increased spectral weight of the non-radiative band-to-band transition at defect B or defect D may simply reduce the number of pumped excitons and cause the limited quantum emission yield.

Consequently, the emission properties should be ascribed to the balance of all contributions described above. For instance, defect C has only negative contributions to decline the emission properties and can be assigned as a complete quenching site. Defect B has a small pre-peak which may contribute to increase radiative decays but might be affected by the Fermi level shift and the increased spectral weight of the non-radiative band-to-band transition. On the other hand, defect D can work as an exciton trapping site and also emit lights from the additionally formed state which is fully consistent with the previously reported red-shifted strong emissions[11,13], though the further discussion should be addressed to conclude the origin of the extra peak.

These findings are quite useful to design the SWNT based photonic devices. In order to improve the efficiency of solar cell, for instance, one should have a long exciton lifetime to earn enough time to separate electrons and holes to create a current. In this viewpoint, defect D must be avoided. Conversely, the single photon emitting device should not consist of emission quenching sites such as defect C.

Furthermore the method described here will find the wide application fields to assign the defect-related excitonic behaviours for other materials such as the diamond or hexagonal boron-nitride which reportedly show the strong emission at defects. It may pave a way to solve common problems for LED (light-emission diode) devices. For example Er doped GaN (blue diodes) has actually many emission lines due to the polymorphic defects and tens of defect structures have been so far proposed, however, both cannot be fully correlated experimentally because single defect emission has never been measured by now. The high-energy resolution EELS will be able to probe single defects for any suitable specimen with diluted defects.

**Acknowledgements:**

This work was supported by KAKENHI (17H04797, 16H06333 and 25220602). T.P. thanks the FWF P27769-N20 for funding.

**Author contributions**

RS and KS designed experiments. YY, TT and HK prepared materials and contributed OAS. RS performed EELS and microscopy. RS and TP analysed data. RS, TP, HK and KS co-wrote the paper. All commented on manuscript.

**Additional information**

The authors declare no competing financial interests. Supplementary information accompanies this paper at xxx.

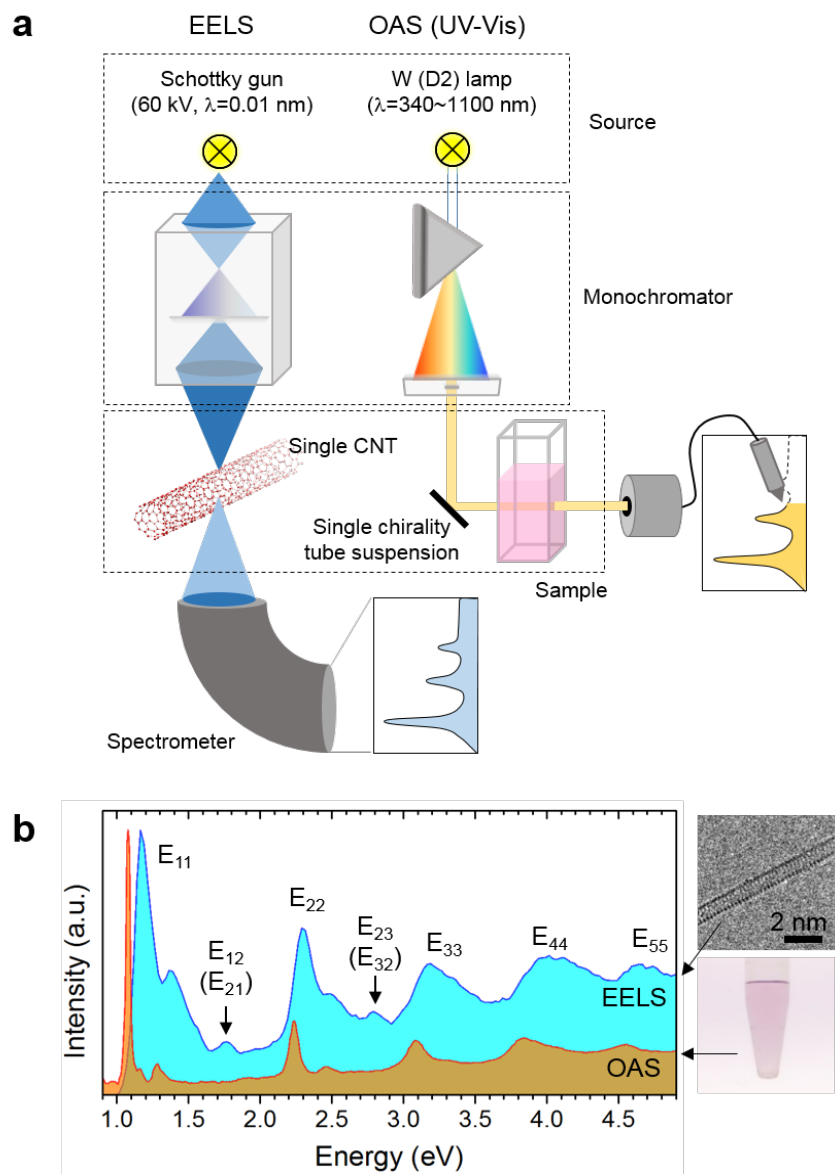

**Figure 1: Schematics of optical absorption spectroscopy (OAS) and electron energy-loss spectroscopy (EELS).** The OA spectrum in **a** was obtained from the chirality separated (9, 2) SWNTs dispersed suspension (the right bottom panel in **b**), while the EEL spectrum in **b** was collected from the single isolated (9, 2) SWNT (the right upper panel in **b**). The chiral index of the SWNT is assigned from the FFT pattern (not shown).

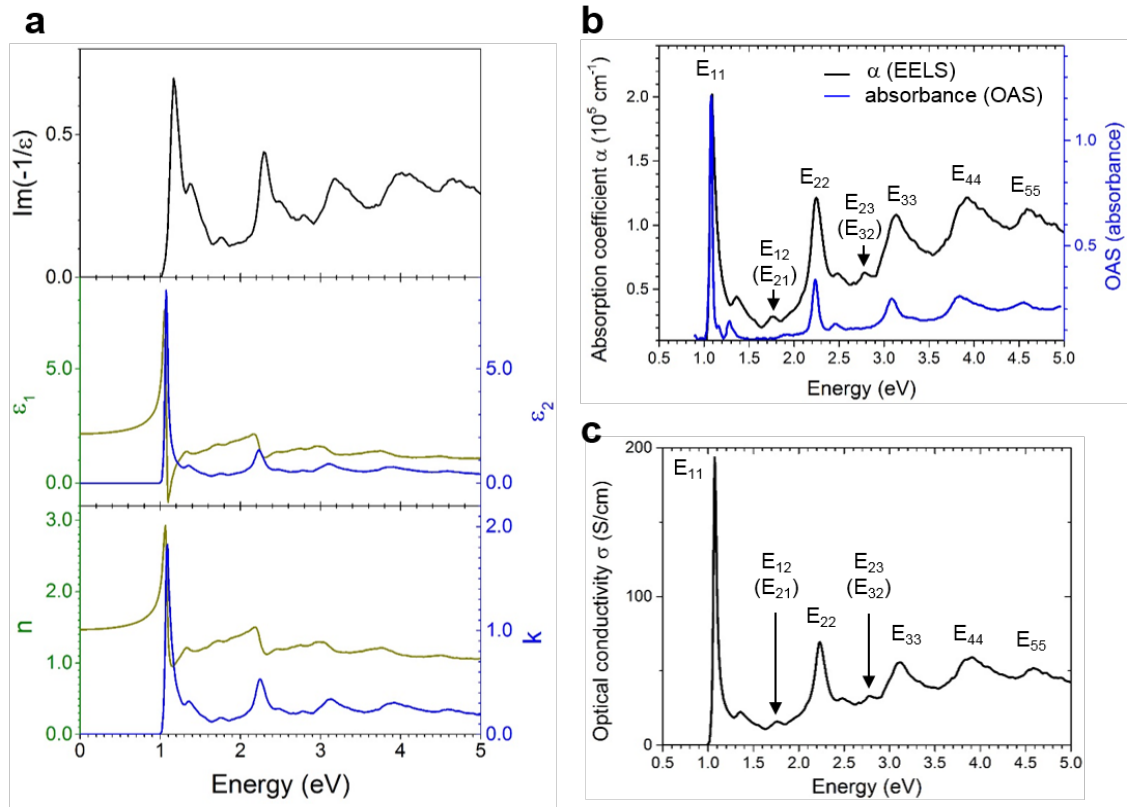

**Figure 2: Experimentally derived optical constants for a single (9, 2) SWNT.** The loss function (the upper panel in **a**) and the dielectric function ($\varepsilon_1$ and $\varepsilon_2$ in the middle panel in **a**) were derived by Kramers-Kronig transformation. The refractive index n and extinction coefficient k (the bottom panel in **a**) were also obtained from the resultant dielectric function. The absorption coefficient $\alpha$ and optical conductivity $\sigma$ (the black lines in **b** and **c**) are also calculated form the complex dielectric functions. The absorption coefficient $\alpha$ is directly compared to the OA spectrum (the blue line in **b**). The broadened OA spectrum with 30 meV Gaussian is also shown in Fig. S2.

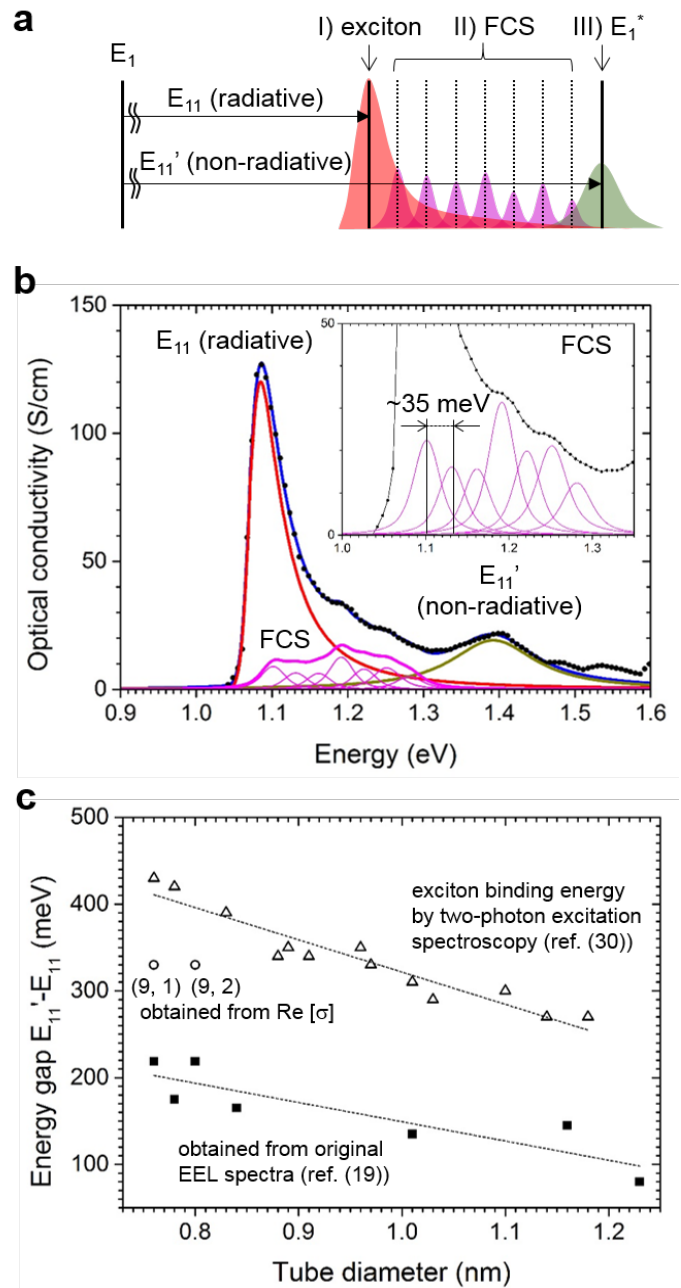

**Figure 3: Fine structure of the lowest excited state. a**, Observable states near the lowest (first) peak in the valence loss consisting of three components: I) an exciton ($E_{11}$) which is asymmetrically modified by the joined DOS of the SWNTs, II) an exciton-phonon coupling (Franck Condon satellites, FCS) and III) a non-radiative band-to-band transition ($E_{11}$'). **b**, Experimentally obtained optical conductivity showing the fine structure near the $E_{11}$ transition (black dots). The initial EEL spectrum was collected from a defect-free region as shown in Fig. 4a. The line shape analysis was performed on the optical conductivity with a reciprocal square root singularity

multiplied by a Gaussian cumulative and a Lorentzian (red line) to simulate the $E_{11}$ exciton peak. Equidistantly separated Voigtians (purple lines) and an additional Voigtian (green line) are accompanied to fit damped FCSs and the non-radiative band to band transition ($E_{11}$'), respectively. All components take into account the experimental broadening by a fixed Gaussian width at 30 meV. **c**, The energy gap between the $E_{11}$ and $E_{11}$' corresponding to the tube diameter. The gaps for (9, 1) and (9, 2) are measured from the optical conductivity (the opened circles). Those for the other chirality nanotubes are measured from unprocessed original EEL spectra (the closed squares). The exciton binding energies measured by two-photon excitation spectroscopy in ref. 30 are also plotted (the opened triangles).

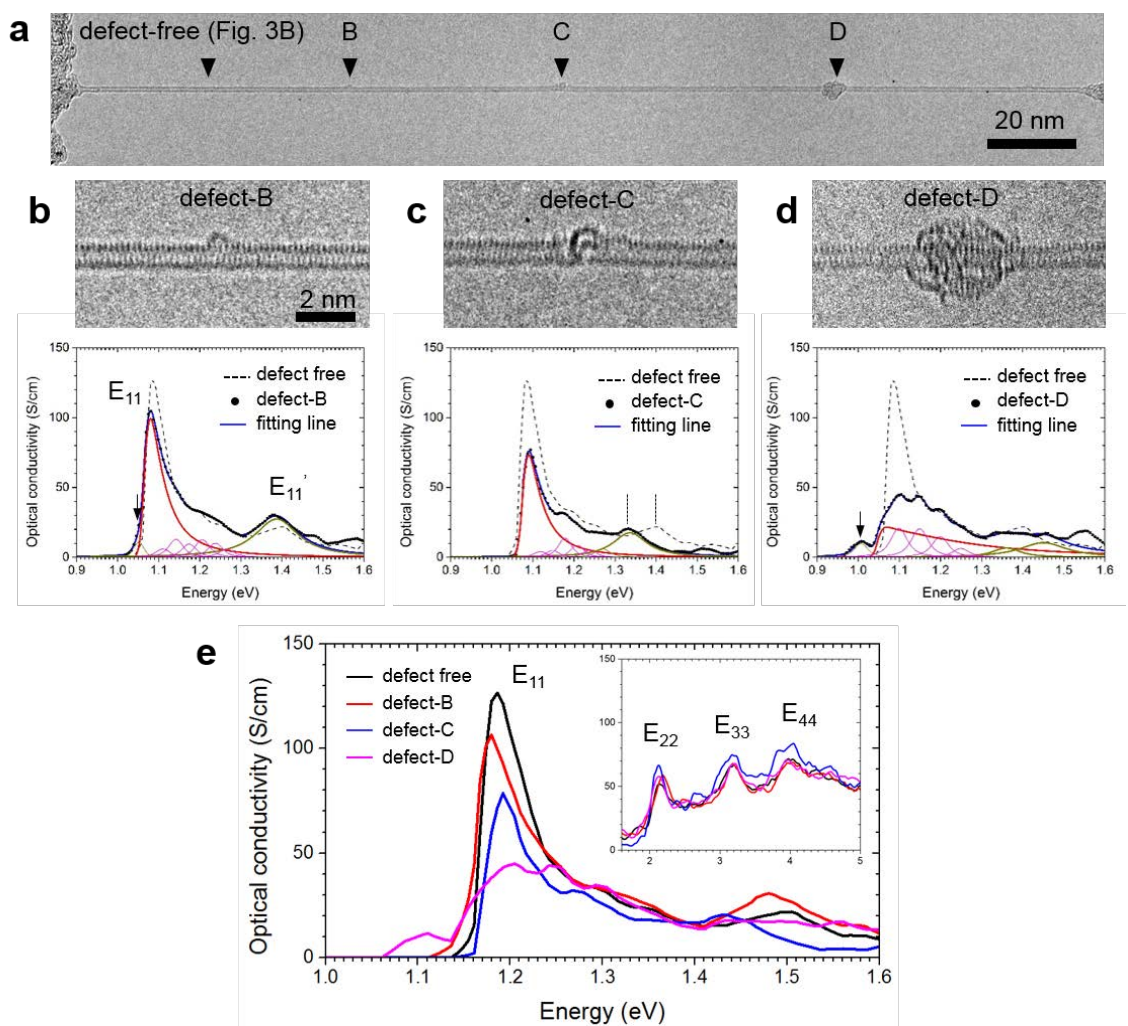

**Figure 4: Modulation of the excitonic behaviours corresponding to the defect types.**
**a**, TEM image for a ~ 230 nm long (9, 2) SWNT suspended between TEM grid with three different defects (namely, defect B, C and D). **b**-**d**, Enlarged TEM images (upper panels) and the optical conductivities (bottom panels) for the each defect plotted by black dots. The case for the defect-free as shown in Fig. 3b is also plotted by a black broken line as a reference in each bottom panel. All the cases are summarised as stacked lines in **e** ($E_{22}$~$E_{44}$ are shown in the inset). The initial EEL spectra were collected when the electron beam scanned on the defects by STEM mode and then transferred to the optical conductivities. The line shape analysis on the optical conductivity for the each defect was performed in the same manner as shown in Fig. 3. The line width of the exciton peak can be extracted from the Lorentzian component in the simulated scaled DOS (listed in Table 1). Only defect D shows an extreme shortening of exciton lifetime.

| | Linewidth of exciton in FWHM (meV) (lifetime) | Integrated intensity | | |
|---|---|---|---|---|
| | | $I_{ex}$ ($I_{ex}/I_{ex,0}$) | $I_{FCS}$ ($I_{FCS}/I_{FCS,0}$) | $I_{nr}$ ($I_{nr}/I_{nr,0}$) |
| defect-free | 47 (340 fs) | 8.88 (1) | 2.91 (1) | 3.60 (1) |
| defect-B | 50 (320 fs) | 7.67 (0.86) | 3.31 (1.14) | 4.87 (1.35) |
| defect-C | 48 (332 fs) | 5.46 (0.61) | 2.25 (0.77) | 2.49 (0.69) |
| defect-D | 283 (57 fs) | 5.50 (0.62) | 4.29 (1.47) | 2.71 (0.75) |

**Table 1 Extracted line width of exciton peak and integrated intensity for each component corresponding to the defect-types.** The line widths are extracted from Lorentzian components in the fitting lines for the exciton peaks as shown in Fig. 3 and 4. The corresponding exciton lifetimes estimated from the simple oscillator model (Fig. S6) are shown in the bottom brackets. Integrated intensity for each component are expressed by $I_{ex}$ (component I), $I_{FCS}$ (component II) and $I_{nr}$ (component III). The relative intensity compared to the defect-free region ($I_{ex,0}$, $I_{FCS,0}$, $I_{nr,0}$) are also shown in the bottom brackets.

# Supplementary information

# Experimental determination of the local optical conductivity of a semiconducting carbon nanotube and its modification at individual defects

**Authors:** Ryosuke Senga[1], Thomas Pichler[2], Yohei Yomogida[1], Takeshi Tanaka[1], Hiromichi Kataura[1] and Kazu Suenaga[1]

**Affiliations:**

[1]*Nano-Materials Research Institute, National Institute of Advanced Industrial Science and Technology (AIST), Tsukuba 305-8565, Japan.*

[2]*Faculty of Physics, University of Vienna, Strudlhofgasse 4, A-1090 Vienna, Austria*

*Correspondence to: Ryosuke Senga (Ryosuke-senga@aist.go.jp) and

Kazu Suenaga (suenaga-kazu@aist.go.jp)

**This PDF file includes:**

    Materials and Methods

    Supplementary text

    Figures S1 to S6

**Materials and Methods:**

Electron microscope and EELS

TEM and STEM-EELS experiments were performed by a JEOL TEM (3C2) equipped with a Schottky field emission gun, a double Wien filter monochromator and delta correctors at 60 keV. EEL spectra were collected by STEM mode in which the energy resolution was set to 30 meV in FWHM. The convergence semiangle $\alpha$ and the EELS collection semiangle $\beta$ were 40 mrad and 70 mrad, respectively. A far-field condition described in Fig. S2 were performed at 10 mrad for both $\alpha$ and $\beta$. The probe current was 10 pA for the valence loss spectroscopy. EELS energy calibration was done by the dual EELS mode which eliminates any mis-calibration due to the zero-loss shift.

Optical absorption spectroscopy

Aqueous solutions of single-wall carbon nanotubes were filled in a quartz glass cell of 10 mm optical path length. Optical absorption spectra were measured using UV-VIS-NIR spectrophotometer (UV-3600, Shimadzu, Kyoto, Japan). In this measurement, to cancel the effect of surfactant, aqueous solution of the surfactant used for the nanotube solution was filled in the other optical cell and was put in the reference light path.

Sample preparation

(9, 2) and (9, 1) single-chirality nanotubes were obtained by a gel column chromatography. Single-wall carbon nanotube (HiPco, NanoIntegris, USA) was dispersed in 0.5 wt% aqueous solution of sodium cholate (98%, Tokyo Chemical Industry, Tokyo, Japan) using ultrasonic homogenizer (Sonifier 250D, Branson Japan) with output 30% for 3 hours. After an ultracentrifugation for 2 hours, upper 80% of supernatant was collected for separation. Sodium dodecyl sulfate (99%, Wako chemical, Japan) was added to be 0.5 wt% in total concentration as second surfactant. Sephacryl (S-200HR, GE Healthcare) was filled in a column as column medium. After loading nanotube solution into the column, a stepwise elution chromatography was conducted to elute out single-chirality nanotubes by adding deoxy sodium cholate (96%, Wako Chemical, Japan) as third surfactant (*22*). In this procedure, for the precise control of the surfactant concentration and separation procedure, high performance liquid chromatography (ÄKTA pure25, GE Healthcare) was used. From many fractions collected in this process, single chirality nanotube solution of (9, 2) and (9, 1) were carefully selected. The purity of each chirality was estimated to be higher than 95% from the optical absorption spectroscopic analysis. The sample for TEM is prepared by dropping the solution on TEM grid and heating at 500C for 5h in the vacuum to remove the surfactant.

**Supplementary text**

Processing of EEL spectra for KK transformation

A KK transformation requires a single scattering EEL spectrum from zero to infinite energy. In the high frequency limit our EEL spectra do not contain any additional peaks related to multiple scatterings (Fig. S3) as it is expected for a single quantum object. Hence we observe a standard $1/\omega^3$ decay and no features around about 45 eV and a further process to deconvolute the multiple scatterings was not needed.

Then, we can just simply have to subtract elastic and quasi-elastic scattering components in the low energy limit. However a zero-loss peak obtained by a monochromated electron source is not straightforwardly fitted by a Lorentzian (or a Lorentzian and a Gaussian) because of its asymmetric shape affected by the vibrational loss around infrared region presenting a quasi-elastic scattering background. Then we have used a combination of a power-law and a Lorentzian functions to fit the tail of the zero loss peak (Fig. S4).

Line shape analysis by a standard oscillator model

From the absolute intensity and linewidths of the exciton peaks in the optical conductivity Re[σ], one can estimate the oscillator strength and the excitons lifetime by fitting it. The standard oscillator model (Kramers-Heisenberg dielectric function) allows a first glance to study the influences of defects on the optical spectra and especially the exciton life times. The model dielectric function is:

$$\varepsilon = \varepsilon_\infty + \sum_j \left( \frac{\omega_{pj}^2(\omega_{Tj}^2 - \omega^2)}{(\omega_{Tj}^2 - \omega^2)^2 + \omega^2\gamma_j^2} + i\frac{\omega\omega_{pj}^2\gamma_j}{(\omega_{Tj}^2 - \omega^2)^2 + \omega^2\gamma_j^2} \right),$$

where $\omega_T$, $\omega_p$ and $\gamma$ denote the transverse frequency, the plasma frequency and the relaxation (damping) factor. The exciton lifetime $\tau$ is simply assumed to be $1/\gamma$ in these model. In order to exclude the experimental broadening effect, the Kramers-Heisenberg dielectric function broadened by a Gaussian with 30 meV width in FWHM is reasonably used for the following analysis to extract the relative changes of the $E_{11}$ peak in the optical conductivity and to assign $\omega_p$ and $\tau$ from the fitting parameters.

Figure S6 shows an example of analysis using this simple oscillator model. The spectrum indicated by the black line in Fig S6a was taken from a defect-free region of (9, 2) SWNT, while the other (the red line) was collected when the electron beam scanned exactly on a defect (the inset in Fig. S6a, identified as defect-A henceforth). The $E_{11}$ peak at the defect clearly shows a decreased intensity as compared to that for the defect-free region (Fig. S6b). On the other hands, the higher order peaks ($E_{22}$~) at defect-A are as almost same as those for the defect-free region (Fig. S6a). The fine structures near the lowest peak for both spectra are fitted by Kramers-Heisenberg dielectric function (Figs. S6c and S6d). Then, a drastic change can be seen in the line width of the main exciton peak for defect-A. From the fitting parameters, the exciton lifetime at defect-A is estimated as 130 fs which decreases by more than 60 % from that for the defect-free region (340 fs). This simply means that the exciton recombination happens more frequently at defect-A. However this simple oscillator model cannot fully fit the asymmetric shape of the exciton peak.

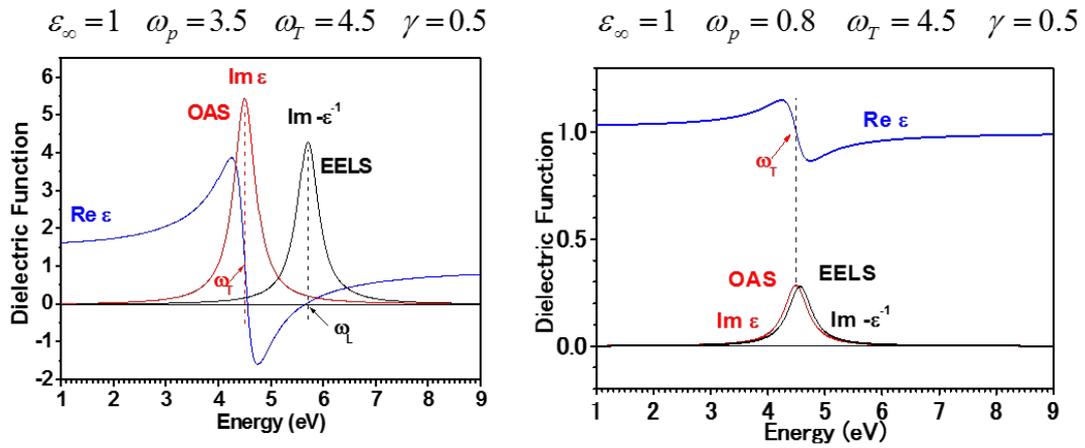

**Figure S1: Model dielectric functions for a bulk and diluted system.** In an oscillator model, a complex dielectric function (Kramers-Heisenberg dielectric function) can be expressed by using the plasma frequency ($\omega_p$), the transvers frequency ($\omega_T$) and the life time (broadening) incident ($\gamma$) as described in Supplementary text. The peak positions in $\varepsilon_2$ (OAS) and $\varepsilon_2/(\varepsilon_1^2+\varepsilon_2^2)$ (EELS) vary depending on the plasma frequency and basically show close values in a diluted system (small $\omega_p$) as shown in the light panel.

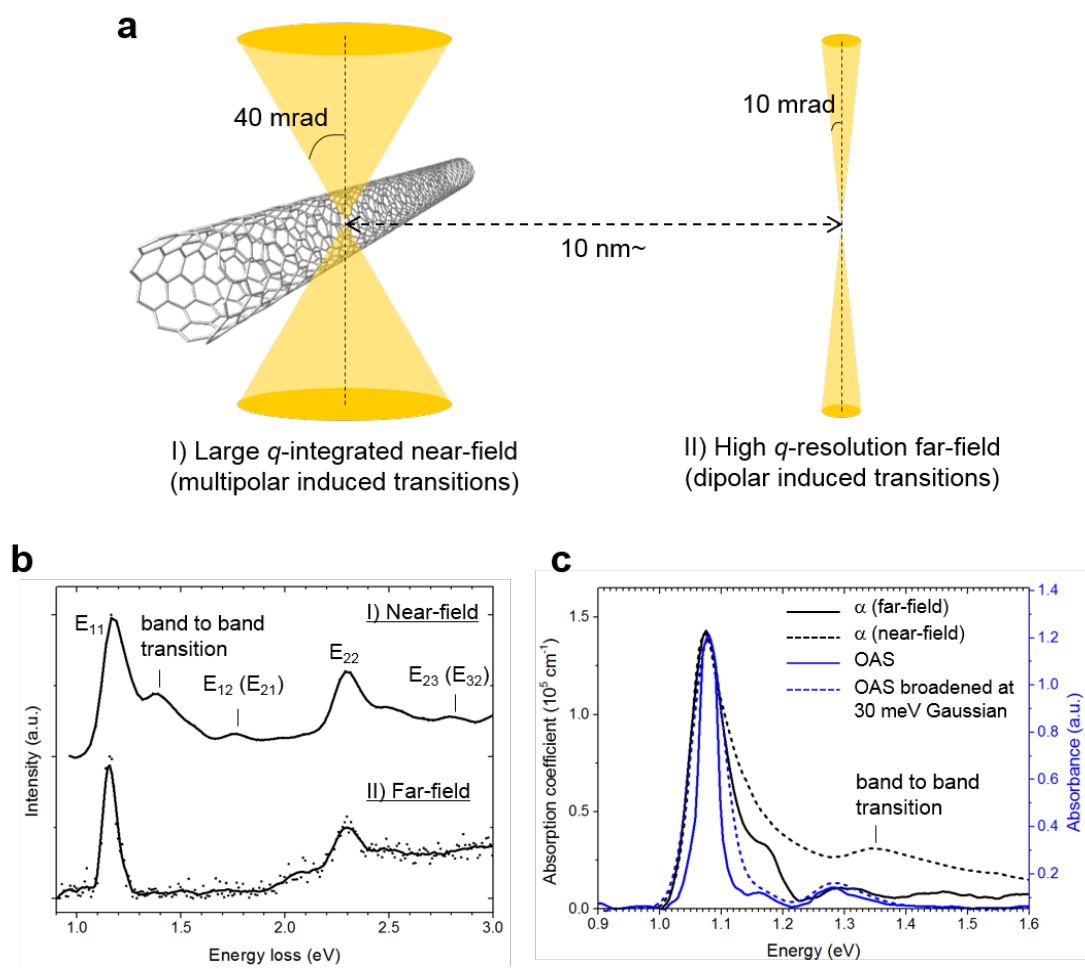

**Figure S2: Different selection rules corresponding to the EELS experimental conditions.** When the electron beam passes on the specimen with a large *q*-momentum space integrated (the left picture in **a**), multipolar induced transitions are allowed (upper EEL spectrum in **b**). On the other hand, when the high *q*-resolution electron beam passes at an aloof position (~ 10nm) apart from the specimen (aloof geometry: the right picture in **a**), only dipolar induced transitions are allowed (the bottom EEL spectrum in **b**) in which the band to band and $E_{ij}$ ($i \neq j$) transitions are silent. Note that the S/N ratio is considerably low at the far-field condition due to the extremely smaller cross-section. This far-field EELS condition provides fully consistent results with optical absorption spectroscopy (the black solid line in **c**).

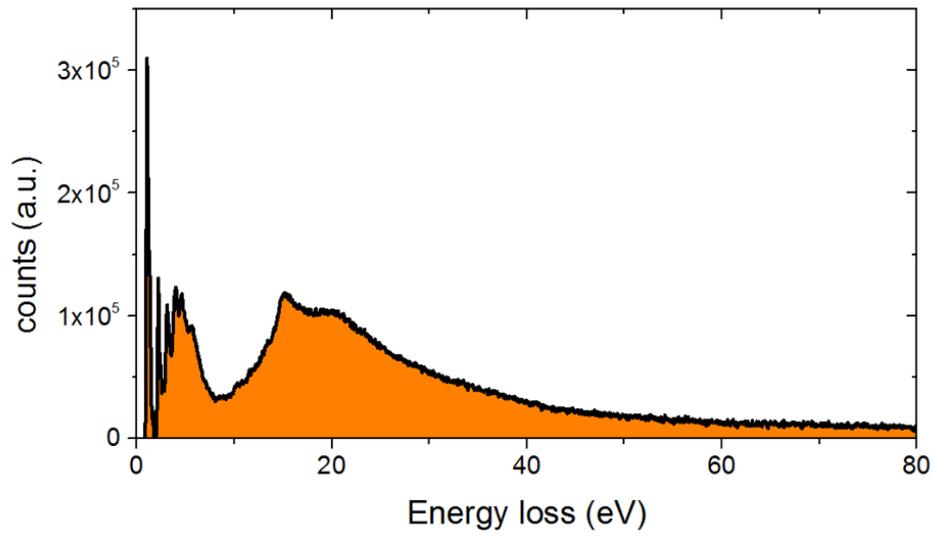

**Figure S3: The full spectrum of a (9, 2) nanotube.** The original EEL spectrum taken from a single isolated (9, 2) nanotube never show any additional peaks related to multiple scatterings

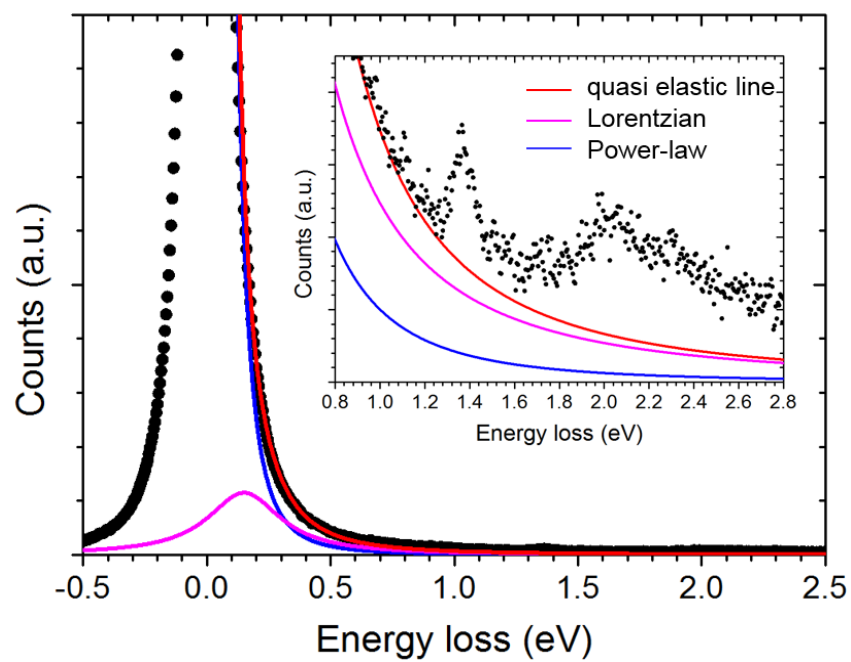

**Figure S4: The background subtraction for a valence loss obtained by the monochromated electron source.** The quasi-elastic line (the red line) is formed by adding the Power-law (the blue line) and the Lorentzian (the purple line) located at around 150 meV which corresponds to the energy-loss for the infrared region.

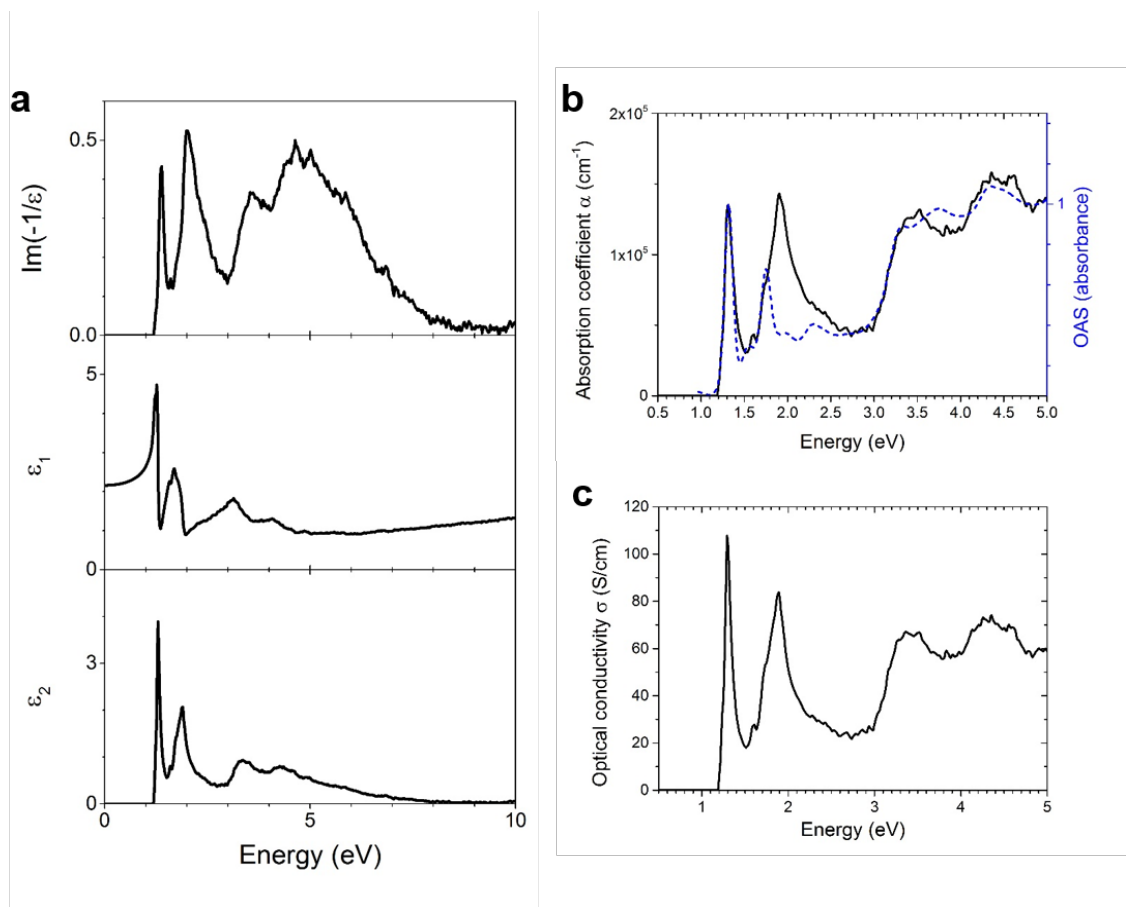

**Figure S5: The optical constants for (9, 1) nanotube derived by Kramers-Kronig analysis.** The loss function (the upper panel in **a**) and the dielectric function ($\varepsilon_1$ and $\varepsilon_2$ in the middle and bottom panels in **a**, respectively) were derived by Kramers-Kronig transformation. The absorption coefficient $\alpha$ and optical conductivity $\sigma$ (the black lines in **b** and **c**) are also calculated form the complex dielectric functions. The absorption coefficient shows a good agreement with the broadened optical absorption spectrum taken from (9, 1) suspension (the blue broken line in **a**) except for the peak around 1.9 eV. This peak can be related to the multipolar-induced transitions which is basically silent in OAS.

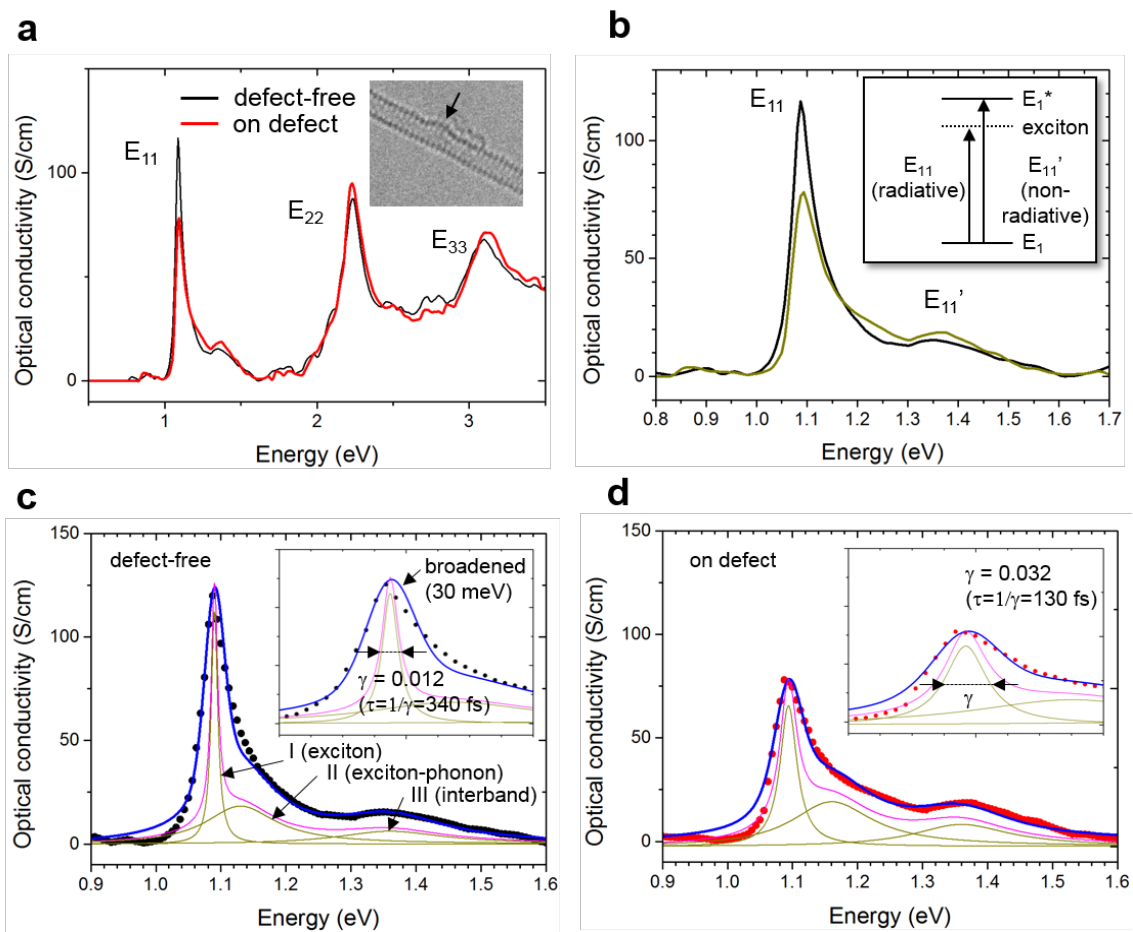

**Figure S6: Line shape analysis by a standard oscillator model.** The optical conductivity obtained from the defective site (the red line in **a** and **b**) and the defect-free region (the black line in **a** and **b**). **c,d**, The line shape analysis by using a standard oscillator model for the defect-free region and the defective site, respectively. The details are described in Supplementary text.